# Non-Majorana Origin of the Half-Quantized Conductance Plateau in Quantum Anomalous Hall Insulator and Superconductor Hybrid Structures

Morteza Kayyalha[1,4], Di Xiao[1,4], Ruoxi Zhang[1,4], Jaeho Shin[1], Jue Jiang[1], Fei Wang[1], Yi-Fan Zhao[1], Ling Zhang[1], Kajetan M. Fijalkowski[2,3], Pankaj Mandal[2,3], Martin Winnerlein[2,3], Charles Gould[2,3], Qi Li[1], Laurens W. Molenkamp[2,3], Moses H. W. Chan[1], Nitin Samarth[1], and Cui-Zu Chang[1]

[1]Department of Physics, The Pennsylvania State University, University Park, PA 16802, USA

[2]Faculty for Physics and Astronomy (EP3), University of Würzburg, Am Hubland, D-97074, Würzburg, Germany

[3]Institute for Topological Insulators, Am Hubland, D-97074, Würzburg, Germany

[4]These authors contributed equally to this work.

Corresponding authors: mhc2@psu.edu (M.H.W.C.); nxs16@psu.edu (N.S.); cxc955@psu.edu (C.Z.C.)

**A quantum anomalous Hall (QAH) insulator coupled to an *s*-wave superconductor is predicted to harbor a topological superconducting phase, the elementary excitations of which (i.e. Majorana fermions) can form topological qubits upon non-Abelian braiding operations. A recent transport experiment interprets the half-quantized two-terminal conductance plateau as the presence of chiral Majorana fermions in a millimeter-size QAH-Nb hybrid structure. However, there are concerns about this interpretation because non-Majorana mechanisms can also generate similar signatures, especially in a disordered QAH system. Here, we fabricated QAH-Nb hybrid structures and studied the QAH-Nb contact transparency and its effect on the corresponding two-terminal conductance. When the QAH** **film is tuned to the metallic regime by electric gating, we observed a sharp zero-bias enhancement in the differential conductance, up to 80% at zero magnetic field. This large enhancement suggests high probability of Andreev reflection and transparent interface between the magnetic topological insulator (TI) and Nb layers. When the magnetic TI film is in the QAH state with well-aligned magnetization, we found that the two-terminal conductance is always half-quantized. Our experiment provides a comprehensive understanding of the superconducting proximity effect observed in QAH-superconductor hybrid structures and shows that the half-quantized conductance plateau is unlikely to be induced by chiral Majorana fermions.**

Topological superconductors (TSCs) are a novel class of quantum matter that host Majorana fermions, particles that are their own anti-particles [1-5]. These Majorana fermions obey non-Abelian statistics and are promising candidates to form a topological qubit, which is the basis for fault-tolerant topological quantum computation [6-8]. TSCs are predicted to appear in a variety of novel condensed matter quantum systems including strong spin-orbit coupled semiconductor-superconductor (SC) hybrid devices [9,10], fractional quantum Hall (QH) systems at filling factor $v = 5/2$ [11,12], spinless $p_x+ip_y$ SCs such as $Sr_2RuO_4$ [2,13], hybrid topological insulator (TI)-SC devices [9], integer QH insulators covered by a conventional $s$-wave SC [14], and thin films of transition metal dichalcogenides [15,16]. Recent theoretical work has predicted a chiral TSC phase when a quantum anomalous Hall (QAH) insulator, a zero magnetic field manifestation of the integer QH effect [17,18], is coupled to an $s$-wave SC [14,19].

The QAH effect has been experimentally demonstrated in thin films of magnetically doped TI [18,20-22]. Ref. [23] recently reported a half-quantized plateau in the two-terminal conductance $\sigma_{1,2}$ measured across a millimeter-size QAH/Nb hybrid structure and interpreted the half-quantized $\sigma_{1,2}$ plateau during magnetization reversal as a "distinct signature" of 1D chiral Majorana edge modes (CMEMs) [19]. Alternative interpretations, however, are also possible. For example, Refs. [24] and [25] theoretically discussed two different scenarios in which a $\sigma_{1,2} = 0.5e^2/h$ plateau can arise without invoking the Majorana physics. Ref. [24] considers the percolation of QAH edges induced by magnetic disorder in the QAH insulator as an alternative origin for the $\sigma_{1,2} = 0.5e^2/h$ plateau. Ref. [25] argues the $\sigma_{1,2} = 0.5e^2/h$ plateau can arise if the SC layer provides good electrical contact to the chiral edge modes of the QAH insulator. In other words, the local equilibrium between the chiral edge modes of the QAH insulators and the SC strip ensures that the total resistance is the series resistance between two QAH sections, each with $h/e^2$ resistance [26].

In this *Letter*, we experimentally studied the effect of the contact transparency in the appearance of the $\sigma_{1,2} = 0.5e^2/h$ plateau. To this end, we fabricated magnetic TI-SC hybrid devices, an example of which is shown in **Figs. 1a** and **1b**. Our device consists of a superconducting Nb strip (width ~ 20 μm) covering the entire width of the QAH layer on the left, a configuration similar to that in Ref. [23], and an additional Nb finger (width ~ 200 nm) on the right (**Figs. 1a** and **1b**). The idea behind our device is two-fold: (*i*) the contact transparency between the magnetic TI and SC layers can be characterized using a differential conductance measurement on the QAH-Nb finger junction [27]; (*ii*) the possible existence of the CMEMs can be investigated by analyzing the two-terminal conductance $\sigma_{1,2}$ measured across the QAH-Nb strip device [19,23]. Furthermore, our QAH film (i.e. magnetic TI) can be tuned to the metallic state using the back-gate voltage $V_g$. This allows us to probe the underlying Andreev reflection mechanisms involved in the magnetic TI-SC hybrid device through the entire phase diagram, i.e., as a function of the chemical potential (tuned by $V_g$) and the external magnetic field. When the QAH layer is tuned to the metallic phase, we observed a strong enhancement of the zero-bias electrical conductance, nearly twice (~180%) the normal-state conductance presumably induced by Andreev reflection. The observation of Andreev reflection in our junction is strong evidence for the induced superconducting pair potential in the magnetic TI layer and allows us to study the effect of transparent interface on the two-terminal conductance $\sigma_{1,2}$ in the QAH-SC hybrid structure. When the magnetic TI is in the QAH regime, we observed that the differential conductance is dominated by the density of state modulation (i.e. breakdown) of the QAH effect. In our experiment, when the QAH and SC layers are strongly coupled as demonstrated by our differential conductance data, $\sigma_{1,2}$ is always half-quantized when the magnetization is well-aligned. We note that our conclusions are supported by measurements on ~ 30 devices fabricated from several sample growths (see Supplementary Materials [28] for more data).

Our Cr-doped (Bi, Sb)$_2$Te$_3$/(Bi, Sb)$_2$Te$_3$/Cr-doped (Bi, Sb)$_2$Te$_3$ QAH sandwich sample is grown on a heat-treated SrTiO$_3$ (111) substrate in a molecular beam epitaxy (MBE) chamber with base pressure below $2 \times 10^{-10}$ mbar. The bottom and top Cr-doped (Bi, Sb)$_2$Te$_3$ layers are 3 quintuple-layer (QL) thick, while the spacer layer is 5 QL (Bi, Sb)$_2$Te$_3$. The Bi/Sb ratio is carefully adjusted during the growth of the sample to ensure the chemical potential is as close to the charge neutral point (CNP) in each layer as possible. The high dielectric constant of SrTiO$_3$ at low temperatures allows us to tune quite effectively by electric gating the chemical potentials within the magnetic exchange gap on both top and bottom surfaces of the QAH sandwich sample [29]. The superconducting strip and finger are patterned using an e-beam lithography followed by the deposition of Nb/Au (100 nm / 5 nm) in a dc sputtering chamber. The normal electrodes are then patterned in the e-beam lithography system and Ti/Au (10 nm / 50 nm) is deposited. Finally, the magnetic TI layer is patterned into a Hall bar geometry and then etched using an Ar plasma. A Physical Property Measurement System (Quantum Design, 2 K, 9 T) cryostat is used to study the magnetic TI-Nb finger contact transparency at 2 K. The measurements on QAH-Nb devices below 2 K are carried out in a dilution refrigerator (Leiden Cryogenics, 10 mK, 9 T).

**Figure 1c** shows the temperature dependence of the Nb finger and the Nb strip resistance. The Nb finger becomes superconducting below its critical temperature $T_{c,\,finger}$ ~ 5 K. The critical temperature of the Nb strip $T_{c,\,strip}$ is ~ 8.6 K. Since we are using a two-terminal technique to measure the resistance, the normal leads contribute ~ 40 Ω to the total resistance, which has been subtracted. **Figure 1d** plots the magnetic field ($\mu_0 H$) dependence of the resistance of the Nb finger and the Nb strip. The Nb strip has an upper critical field $\mu_0 H_{c2,\,strip}$ ~ 2.9 T. **Figures 1e** and **1f** show the $\mu_0 H$ dependence of the longitudinal resistance (conductance) $\rho_{xx}$ ($\sigma_{xx}$) and the Hall resistance (conductance) $\rho_{yx}$ ($\sigma_{xy}$) at $V_g = V_g^0 = +42$ V and $T = 30$ mK, where typical QAH characteristics, quantized $\rho_{yx}$ ($\sigma_{xy}$) accompanied by vanishing $\rho_{xx}$ ($\sigma_{xx}$), are observed. Since the $\rho_{xx}$ peak value during magnetization reversal is comparable to the quantized $\rho_{yx}$ value, the zero Hall conductance $\sigma_{xy} = 0$ plateau (i.e. Chern number $C = 0$ phase [30]) is not observable. The $\sigma_{xy} = 0$ plateau is usually observed in the thinner uniformly-doped QAH samples with a larger $\rho_{xx}$ peak [31,32].

We characterized the interface transparency between the magnetic TI-Nb finger junction by measuring its differential conductance, which is related to the probabilities of the Andreev reflection (AR) and the normal reflection (NR) across the interface. **Figures 2a** and **2b** show the differential upstream conductance $\sigma_U = dI_{6,8}/dV_{7,8}$ and the downstream conductance $\sigma_D = dI_{6,8}/dV_{9,8}$, where the subscript numbers correspond to the electrodes shown in **Fig. 1a**, at different magnetic fields. $\sigma_U$ and $\sigma_D$ are normalized by their respective values at $T > T_{c,\,finger}$ (i.e. $\sigma_{6K}$). For $V_g = V_g^0$, an interplay between AR and NR at the interface as well as the breakdown of the QAH system [33-35] determine the differential conductance, which turns out to be dominated by the breakdown of the QAH state in our samples (see **Fig. S3** in Supplementary Materials [28]). On the other hand, $\sigma_U$ ($\sigma_D$) is a better probe of the AR to NR ratio when the magnetic TI is in its metallic phase, as we will discuss below. In order to characterize the magnetic TI-Nb interface transparency, we applied a negative $V_g = -50$ V to reach the metallic phase of the magnetic TI. At zero magnetic field, we observe an enhancement of the zero-bias conductance, approaching 180% of its high-temperature value, revealing a highly transparent magnetic TI-SC interface. Remarkably, with increasing $\mu_0 H$, (*i*) the superconductivity is suppressed in the Nb finger, consistent with the $\mu_0 H$ dependence of resistance data (**Fig. 1d**) and (*ii*) the zero-bias conductance enhancement remains almost the same, indicating the probability of AR at the interface is unchanged, i.e., the magnetic TI-Nb contact transparency is unaffected by the increasing $\mu_0 H$. For $\mu_0 H$ larger than the coercive field ($\mu_0 H_c \sim 0.06$ T) of the magnetic TI layer at $T = 2$ K, zero-bias $\sigma_U$ ($\sigma_D$) is slightly reduced (increased). This reduction (enhancement) of $\sigma_U$ ($\sigma_D$) is likely a result of the magnetization reversal in the magnetic TI layer around $\mu_0 H_c$ regime (see **Fig. S4** in Supplementary Materials for the magnetoresistance loops of $R_U \sim 1/\sigma_U$ and $R_D \sim 1/\sigma_D$ [28]). In our experiment, it is hard to extract the voltage drop across the magnetic TI-Nb junction because the magnetic TI layer is still very resistive even in its metallic phase (usually several hundred $\Omega$); i.e., most of the voltage drop appears across the resistive part of the magnetic TI layer rather than the magnetic TI-Nb interface. Therefore, we plot the differential conductance as a function of the dc current $I_{dc}$ rather than the dc voltage $V_{dc}$. To confirm the sharp zero-bias conductance peak is indeed a result of the AR process at the interface, we studied the temperature dependence of $\sigma_U$ ($\sigma_D$) vs $I_{dc}$ in **Figs. 2c** and **2d**, where we observed a featureless $\sigma_U$ ($\sigma_D$) at $T = 6$ K $> T_{c, \text{finger}}$. We note that at $T = 6$ K, the Nb finger is no longer superconducting (**Fig. 1c**), and thus the differential conductance is a sum of the contributions from the NR at the interface and the resistive part of the magnetic TI film. Therefore, the zero-bias conductance at $T = 6$ K takes the same value as that of the high-bias regime for $T \leq 5$ K, consistent with the AR picture for normal metal-superconductor junctions [36,37], for more details see Supplemental Materials [28].

Our experimental observations reveal the presence of a highly transparent interface between the magnetic TI and Nb finger throughout the $\mu_0 H$ range of our interest (0 T $< \mu_0 H <$ 1 T). Given the much larger contact area of the Nb strip and the similar fabrication processes undergone by both the Nb finger and strip, we expect the interface transparency to be similar for the magnetic TI-Nb strip as well. Moreover, the transparent interface and the chiral nature of the edge modes in the QAH regime are expected to ensure that an electron propagating along the Nb strip, due to its large size, will quickly become an equal mixture of electrons and holes [27].

Utilizing the transparent contacts, we next focus on the Hall bar structure with the Nb strip covering the entire width of the magnetic TI film. Our QAH-SC hybrid device (minus the Nb finger) shown in **Figs. 1a** and **1b** is similar to the device employed in Ref. [23]. A $\sigma_{1,2} \sim 0.5e^2/h$ plateau during the magnetization reversal ($\sim \mu_0 H_c$) followed by a $\sigma_{1,2} \sim e^2/h$ plateau for $\mu_0 H_c < |\mu_0 H|/< \mu_0 H_{c2, \text{strip}}$ regime is reported in Ref. [23]. They interpreted that (*i*) this behavior is induced by the presence of the CMEMs and (*ii*) the transition from $\sigma_{1,2} = e^2/h$ plateau to $\sigma_{1,2} = 0.5e^2/h$ plateau corresponds to a topological phase transition in the TSC state from $N = 2$ to $N = 1$, where $N$ denotes the number of the CMEMs [14,19]. In the same structure, an extremely small two-terminal conductance $\sigma_{1,3}$, measured between the Nb strip and the QAH sample, for $\mu_0 H_c < |\mu_0 H|/ < \mu_0 H_{c2, \text{strip}}$ is also reported [23]. The small value of $\sigma_{1,3}$ in this $\mu_0 H$ range indicates the Nb layer is likely decoupled from the QAH sample and hence the $\sigma_{1,2} = e^2/h$ plateau may be a result of poor electrical contact between the QAH insulator and the Nb layers, i.e., there is no proximity-induced superconductivity and no AR at the QAH-Nb interface [23,25]. We note that the observation of $\sigma_{1,2} = 0.5e^2/h$ in the QAH insulator phase is not unusual [24,25]. In fact, a normal metal (e.g. gold), overlaying the two edges of the QAH sample will give rise to such a quantization in $\sigma_{1,2}$ [26].

**Figure 3a** displays the $\mu_0 H$ dependence of the two-terminal conductance $\sigma_{1,2}$ for $V_g = V_g^0 = +42$ V of our device. In contrast to Ref. [23], we observed that $\sigma_{1,2} = dI_{13,6}/dV_{1,2} \sim 0.5e^2/h$ over the entire range of the magnetic field except in the $\mu_0 H$ range when the magnetization of the sample is being reversed near $\mu_0 H_c$. In this range, $\sigma_{1,2}$ drops to $\sim 0.21e^2/h$. Specifically, no change in $\sigma_{1,2}$ is observed when $\mu_0 H$ is increased across the critical field value of the Nb strip, i.e. $\mu_0 H_{c2,\text{strip}} \sim 2.9$ T (**Fig. 1d**). We also measured $\sigma_{1,3} = dI_{13,6}/dV_{1,3}$, the conductance between the Nb strip and the QAH sample, as shown in **Fig. 3b**. $\sigma_{1,3}$ is found to be $\sim e^2/h$ in the entire $|\mu_0 H| > \mu_0 H_c$ range, indicating the Nb strip is strongly coupled to the QAH sample, leading to the equilibrium of chemical potentials between chiral edge modes of the QAH sample and bulk Nb layer [25]. This behavior is what one would expect if a normal metal was used instead of the Nb strip. For $\mu_0 H > \mu_0 H_{c2,\text{strip}}$, the Nb strip turns into the normal state, hence, $\sigma_{1,2}$ remains half-quantized. Out of thoroughness, we have also studied 9 QL V-doped TI samples, which were previously demonstrated to exhibit perfect QAH effect [35,38-40] and signatures of axion electrodynamics [38]. The devices were patterned using an optical lithography process and used MoRe as the SC strip. We again observed the $\sigma_{1,2} \sim 0.5e^2/h$ plateau for the entire $\mu_0 H$ region with well-aligned magnetization (see **Fig. S8** in Supplementary Material [28]).

The existence of the zero Hall conductance plateau with the $C = 0$ phase in a QAH sample was claimed as a prerequisite for the observation of the $N = 1$ TSC phase [23]. Citing Ref. [19], the transition from the $C = 0$ (i.e. $N = 0$) phase to the $C = 1$ (i.e. $N = 2$) phase is given in Ref. [23] as the key evidence for the existence of the $N = 1$ TSC phase. We note, however, the theoretical calculations in Ref. [19] treated the superconductor strip merely as the "source" of the small energy gap while overlooking the fact that the strip also serves as an electrical short for the QAH device.

Our results, on the other hand, show that the $\sigma_{1,2} = e^2/h$ plateau in the $C = 1$ phase is very likely a result of decoupling of the QAH insulator from the Nb layer. Hence, it is not predicated upon the existence of a TSC phase with $N = 2$. To exclude the possibility that the $\sigma_{1,2} \sim 0.5e^2/h$ plateau observed in our thick QAH sandwich sample is due to the absence of the zero Hall conductance plateau (i.e. the $C = 0$ phase), we carried out measurements on QAH samples with the $C = 0$ phase. We fabricated two 6QL Cr-doped (Bi, Sb)$_2$Te$_3$ samples, similar to the ones used in Ref. [23]. Next, we scratched both samples into millimeter-size Hall bar structures (0.5 mm × 1 mm) and then sputtered Nb strips onto the samples with a mechanical mask. We measured $\sigma_{1,2}$ across one and two Nb strips, respectively. $\sigma_{1,2}$ for one Nb strip sample (**Fig. 4a**) is similar to that measured in the QAH sandwich sample (**Fig. 3a**). Therefore, the existence or the non-existence of the zero Hall conductance plateau in QAH samples does not change our findings, specifically, the $\sigma_{1,2} \sim 0.5e^2/h$ plateau is observed for the entire $\mu_0H$ region with well-aligned magnetization. We also studied the V-doped TI/TI/Cr-doped TI QAH sandwich samples, in which a well-established $C = 0$ insulating phase (i.e. the axion insulator state) emerges [41,42]. Here, we also observed the $\sigma_{1,2} \sim 0.5e^2/h$ plateau for the entire $\mu_0H$ region with well-aligned magnetization (see **Fig. S7** in Supplementary Material [28]). To better understand the relation between $\sigma_{1,2}$ and the coupling of the SC layer to the chiral edge modes, we measured $\sigma_{1,2}$ across a 6QL Cr-doped (Bi, Sb)$_2$Te$_3$ QAH sample with two Nb strips. $\sigma_{1,2} \sim e^2/3h$ is observed for the entire well-aligned $\mu_0 H$ regimes (**Fig. 4b**). The value of the $\sigma_{1,2}$ plateau decreases with increasing the number of Nb strips (*n*), specifically, $\sigma_{1,2} \sim e^2/(n+1)h$, indicating the total two-terminal resistance $\rho_{1,2}$ is a series resistance of (*n*+1) QAH sections, each contributing $h/e^2$ [26].

In summary, we observed an enhancement of the zero-bias differential conductance in the QAH-SC junction, indicating that Andreev reflection is the dominant mechanism at the interface and the QAH-SC contact is highly transparent. Using such transparent interfaces, we investigated the origin of the half-quantized two terminal plateau measured across a QAH insulator covered by a SC strip. Our experimental results demonstrate that if the SC layer is strongly coupled to the QAH sample, the two-terminal conductance $\sigma_{1,2}$ is half-quantized throughout the magnetic field range, where the magnetization is well-aligned. Perfect agreement between the data obtained from the various QAH samples and sample geometries demonstrate the robustness, reproducibility and universality of the presented phenomena. Therefore, we conclude that the observation of $\sigma_{1,2} \sim 0.5e^2/h$ plateau cannot be considered as evidence for the existence of chiral Majorana edge modes and $N = 1$ TSC phase in the millimeter-size QAH-SC hybrid structures.

**Acknowledgments**

The authors would like to thank C. X. Liu, K.T. Law, B. Lian, J. Wang, X. Dai, J. Jain, H. Z. Lu, Z. Wang, B. H. Yan, G. H. Lee, Y. L. Chen, K. He, W. J. Ji, Q. K. Xue and X. D. Xu for helpful discussions. D. X. and N. S. acknowledge support from ONR grant (N-000141512370) and Penn State 2DCC-MIP under NSF grant (DMR-1539916). M. H. W. C. acknowledge support from NSF grant (DMR-1707340). C. Z. C. acknowledges support from NSF-CAREER award (DMR-1847811) and Alfred P. Sloan Research Fellowship. Support for transport measurements and data analysis at Penn State is provided by DOE grant (DE-SC0019064). C. G and L.W. M. acknowledge support from EU ERC-AG Programs (project 3-TOP and 4TOPS).

# Figures and figure captions

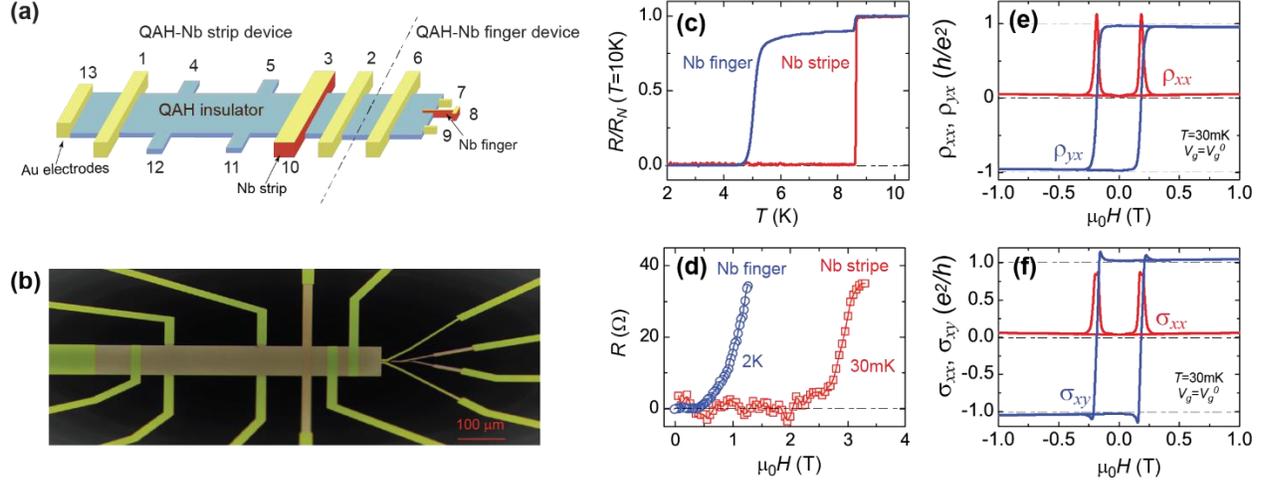

**Figure 1| QAH-Nb devices.** (a) Schematic of the device consisting of a QAH insulator layer, a Nb strip, and a Nb finger. The Nb finger is used to characterize the magnetic TI-Nb contact transparency, while the Nb strip is employed to study the two-terminal conductance $\sigma_{1,2}$ across the QAH-Nb structure. (b) An optical microscope image of the QAH-Nb device. (c) Temperature dependence of the normalized resistance of the Nb finger and Nb strip. (d) $\mu_0H$ dependence of the resistance of the Nb finger and Nb strip. The drop in the resistance of the Nb finger at $T \sim 8.6$ K is associated with a superconducting transition of the Nb section with a larger width ($\sim 4$ μm) in the device, see **Fig.1b** and the inset of **Fig. 2a**. (e, f) The four-terminal longitudinal and Hall resistance ($\rho_{xx}$ and $\rho_{yx}$) (e) and their corresponding longitudinal and Hall conductance ($\sigma_{xx}$ and $\sigma_{xy}$) (e) as a function of $\mu_0H$ measured at $V_g = V_g^0 = +42$V and $T = 30$ mK.

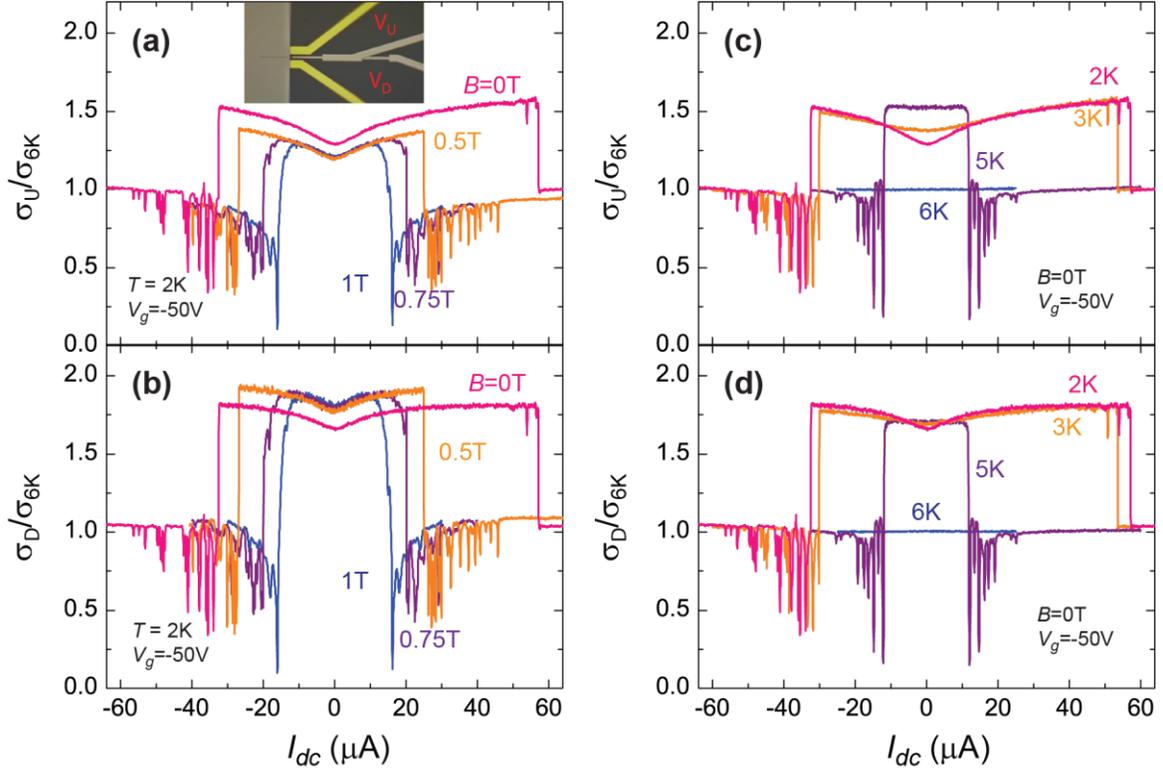

**Figure 2| Contact transparency in the magnetic TI-Nb finger device.** (a, b) The differential upstream conductance $\sigma_U = dI_{6,8}/dV_{7,8}$ (a) and the downstream conductance $\sigma_D = dI_{6,8}/dV_{9,8}$ (b) of the magnetic TI-Nb finger junction normalized by their high temperature ($T > T_{c,\text{finger}}$) values $\sigma_{6K}$, measured at different $\mu_0 H$ and $T = 2$ K. Inset of (a) shows a magnified optical image of the magnetic TI-Nb finger device. (c, d) The normalized $\sigma_U$ (c) and $\sigma_D$ (d) measured at different temperatures and zero magnetic field. An ac excitation current $I_{ac} = 10$ nA is employed for the differential conductance measurements.

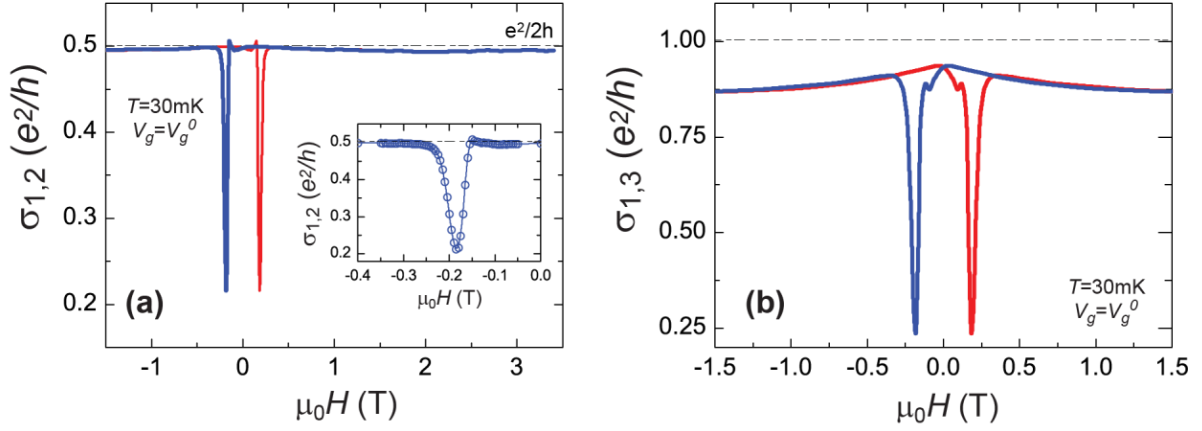

**Figure 3| Two-terminal conductance $\sigma_{1,2}$ across the QAH-Nb strip device.** (a) $\mu_0 H$ dependence of two-terminal conductance $\sigma_{1,2} = dI_{13,6}/dV_{1,2}$ measured at $V_g = V_g^0 = +42$ V and $T = 30$ mK. $\sigma_{1,2} \sim 0.5 e^2/h$ for the entire $\mu_0 H$ range when the magnetization is well-aligned. No change in $\sigma_{1,2}$ is observed when the Nb strip transitions from the superconducting state to the normal state ($\mu_0 H > \mu_0 H_{c2,\,\text{strip}} \sim 2.6$ T). Inset magnifies the $\mu_0 H$ axis during the magnetization reversal process. (b) $\mu_0 H$ dependence of two-terminal conductance $\sigma_{1,3} = dI_{13,6}/dV_{1,3}$. $\sigma_{1,3}$ approaches $\sim e^2/h$ for $|\mu_0 H| > \mu_0 H_c$, indicating good contact transparency between the Nb strip and the QAH sample. An ac excitation current $I_{ac} = 1$ nA is employed for the measurement of $\sigma_{1,2}$ and $\sigma_{1,3}$.

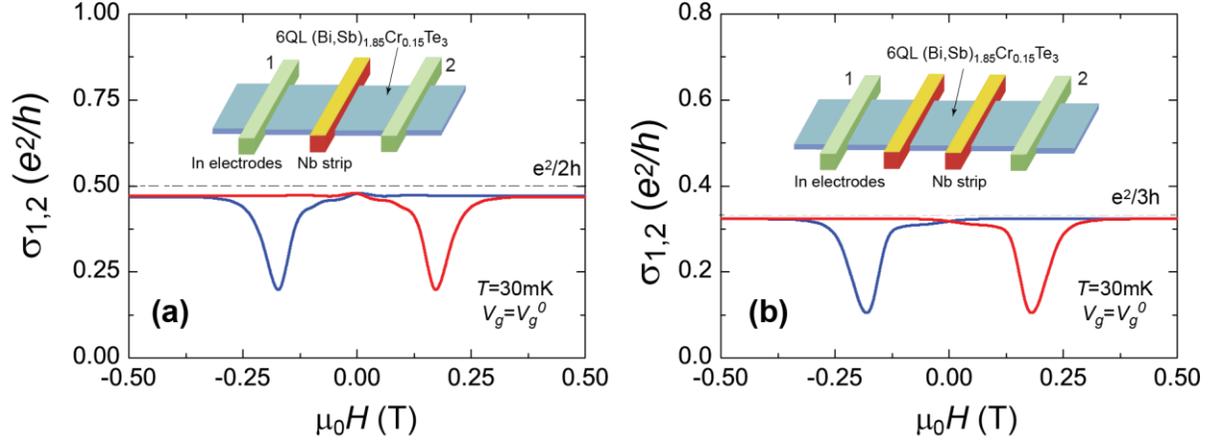

**Figure 4| Two-terminal conductance $\sigma_{1,2}$ in 6QL uniformly doped QAH-Nb strip devices.**
(a,b) $\mu_0 H$ dependence of two-terminal conductance $\sigma_{1,2}$ measured across one Nb strip (a) and two Nb strips stacked on a 6QL (Bi, Sb)$_{1.85}$Cr$_{0.15}$Te$_3$ QAH sample at $V_g = V_g^0$ and $T = 30$ mK. Insets in (a) and (b) show the corresponding device configuration. With increasing the number (*n*) of Nb strips, the corresponding $\sigma_{1,2}$ plateau decreases as $\sigma_{1,2} \sim e^2/(n+1)h$.

# References

[1] E. Majorana, *Nuovo Cim.* **14**, 171 (1937).

[2] N. Read and D. Green, *Phys. Rev. B* **61**, 10267 (2000).

[3] C. W. J. Beenakker, *Annu. Rev. Condens. Matter Phys.* **4**, 113 (2013).

[4] J. Alicea, *Rep. Prog. Phys.* **75**, 076501 (2012).

[5] X. L. Qi and S. C. Zhang, *Rev. Mod. Phys.* **83**, 1057 (2011).

[6] A. Y. Kitaev, *Ann. Phys.* **303**, 2 (2003).

[7] F. Wilczek, *Nat. Phys.* **5**, 614 (2009).

[8] C. Nayak, S. H. Simon, A. Stern, M. Freedman, and S. Das Sarma, *Rev. Mod. Phys.* **80**, 1083 (2008).

[9] L. Fu and C. L. Kane, *Phys. Rev. Lett.* **100**, 096407 (2008).

[10] R. M. Lutchyn, J. D. Sau, and S. Das Sarma, *Phys. Rev. Lett.* **105**, 077001 (2010).

[11] G. Moore and N. Read, *Nucl. Phys. B* **360**, 362 (1991).

[12] A. Stern, *Nature* **464**, 187 (2010).

[13] A. P. Mackenzie and Y. Maeno, *Rev. Mod. Phys.* **75**, 657 (2003).

[14] X. L. Qi, T. L. Hughes, and S. C. Zhang, *Phys. Rev. B* **82**, 184516 (2010).

[15] N. F. Q. Yuan, K. F. Mak, and K. T. Law, *Phys. Rev. Lett.* **113**, 097001 (2014).

[16] Y. T. Hsu, A. Vaezi, M. H. Fischer, and E. A. Kim, *Nat. Commun.* **8**, 14985 (2017).

[17] F. D. M. Haldane, *Phys. Rev. Lett.* **61**, 2015 (1988).

[18] C. Z. Chang, J. S. Zhang, X. Feng, J. Shen, Z. C. Zhang, M. H. Guo, K. Li, Y. B. Ou, P. Wei, L. L. Wang, Z. Q. Ji, Y. Feng, S. H. Ji, X. Chen, J. F. Jia, X. Dai, Z. Fang, S. C. Zhang, K. He, Y. Y. Wang, L. Lu, X. C. Ma, and Q. K. Xue, *Science* **340**, 167 (2013).

[19] J. Wang, Q. Zhou, B. Lian, and S. C. Zhang, *Phys. Rev. B* **92**, 064520 (2015).

[20] C. Z. Chang, W. W. Zhao, D. Y. Kim, H. J. Zhang, B. A. Assaf, D. Heiman, S. C. Zhang, C. X. Liu, M. H. W. Chan, and J. S. Moodera, *Nat. Mater.* **14**, 473 (2015).

[21] J. G. Checkelsky, R. Yoshimi, A. Tsukazaki, K. S. Takahashi, Y. Kozuka, J. Falson, M. Kawasaki, and Y. Tokura, *Nat. Phys.* **10**, 731 (2014).

[22] X. F. Kou, S. T. Guo, Y. B. Fan, L. Pan, M. R. Lang, Y. Jiang, Q. M. Shao, T. X. Nie, K. Murata, J. S. Tang, Y. Wang, L. He, T. K. Lee, W. L. Lee, and K. L. Wang, *Phys. Rev. Lett.* **113**, 137201 (2014).

[23] Q. L. He, L. Pan, A. L. Stern, E. C. Burks, X. Y. Che, G. Yin, J. Wang, B. Lian, Q. Zhou, E. S. Choi, K. C. Murata, X. F. Kou, Z. J. Chen, T. X. Nie, Q. M. Shao, Y. B. Fan, S. C. Zhang, K. Liu, J. Xia, and K. L. Wang, *Science* **357**, 294 (2017).

[24] Y. Y. Huang, F. Setiawan, and J. D. Sau, *Phys. Rev. B* **97**, 100501 (2018).

[25] W. J. Ji and X. G. Wen, *Phys. Rev. Lett.* **120**, 107002 (2018).

[26] C. Z. Chang, W. W. Zhao, D. Y. Kim, P. Wei, J. K. Jain, C. X. Liu, M. H. W. Chan, and J. S. Moodera, *Phys. Rev. Lett.* **115**, 057206 (2015).

[27] G. H. Lee, K. F. Huang, D. K. Efetov, D. S. Wei, S. Hart, T. Taniguchi, K. Watanabe, A. Yacoby, and P. Kim, *Nat. Phys.* **13**, 693 (2017).

[28] See Supplemental Material at XXXXX for further details regarding the device image, differential conductance at the QAH phase, the axion insulator-Nb strip sample, and the results from Würzburg.

[29] J. Jiang, D. Xiao, F. Wang, J.-H. Shin, D. Andreoli, J. Zhang, R. Xiao, Y.-F. Zhao, M. Kayyalha, L. Zhang, K. Wang, J. Zang, C. Liu, N. Samarth, M. H. W. Chan, and C.-Z. Chang, arXiv:1901.07611 (2019).

[30] J. Wang, B. Lian, and S. C. Zhang, *Phys. Rev. B* **89**, 085106 (2014).

[31] X. F. Kou, L. Pan, J. Wang, Y. B. Fan, E. S. Choi, W. L. Lee, T. X. Nie, K. Murata, Q. M. Shao, S. C. Zhang, and K. L. Wang, *Nat. Commun.* **6**, 8474 (2015).

[32] Y. Feng, X. Feng, Y. B. Ou, J. Wang, C. Liu, L. G. Zhang, D. Y. Zhao, G. Y. Jiang, S. C. Zhang, K. He, X. C. Ma, Q. K. Xue, and Y. Y. Wang, *Phys. Rev. Lett.* **115**, 126801 (2015).

[33] M. Kawamura, R. Yoshimi, A. Tsukazaki, K. S. Takahashi, M. Kawasaki, and Y. Tokura, *Phys. Rev. Lett.* **119**, 016803 (2017).

[34] E. J. Fox, I. T. Rosen, Y. F. Yang, G. R. Jones, R. E. Elmquist, X. F. Kou, L. Pan, K. L. Wang, and D. Goldhaber-Gordon, *Phys. Rev. B* **98**, 075145 (2018).

[35] M. Götz, K. M. Fijalkowski, E. Pesel, M. Hartl, S. Schreyeck, M. Winnerlein, S. Grauer, H. Scherer, K. Brunner, C. Gould, F. J. Ahlers, and L. W. Molenkamp, *Appl. Phys. Lett.* **112**, 072102 (2018).

[36] G. E. Blonder, M. Tinkham, and T. M. Klapwijk, *Phys. Rev. B* **25**, 4515 (1982).

[37] M. Tinkham, *Introduction to superconductivity*, 2nd edn., International series in pure and applied physics.

[38] S. Grauer, K. M. Fijalkowski, S. Schreyeck, M. Winnerlein, K. Brunner, R. Thomale, C. Gould, and L. W. Molenkamp, *Phys. Rev. Lett.* **118**, 246801 (2017).

[39] M. Winnerlein, S. Schreyeck, S. Grauer, S. Rosenberger, K. M. Fijalkowski, C. Gould, K. Brunner, and L. W. Molenkamp, *Phy. Rev. Mater.* **1**, 011201 (2017).

[40] S. Grauer, S. Schreyeck, M. Winnerlein, K. Brunner, C. Gould, and L. W. Molenkamp, *Phys. Rev. B* **92**, 201304 (2015).

[41] D. Xiao, J. Jiang, J. H. Shin, W. B. Wang, F. Wang, Y. F. Zhao, C. X. Liu, W. D. Wu, M. H. W. Chan, N. Samarth, and C. Z. Chang, *Phys. Rev. Lett.* **120**, 056801 (2018).

[42] M. Mogi, M. Kawamura, A. Tsukazaki, R. Yoshimi, K. S. Takahashi, M. Kawasaki, and Y. Tokura, *Sci. Adv.* **3**, eaao1669 (2017).